\documentclass[reprint,amsmath,amssymb,aps,prd,nofootinbib,superscriptaddress]{revtex4-2}

\usepackage{graphicx}
\usepackage{dcolumn}
\usepackage{bm}
\usepackage{hyperref}
\usepackage[mathlines]{lineno}
\usepackage{amsfonts,color,xcolor}

\begin{document}

\title{Lorentz gauge-invariant variables in torsion-based theories of gravity}

\author{Daniel Blixt}
\altaffiliation[]{daleblixt@gmail.com} \affiliation{Laboratory of
Theoretical Physics, Institute of Physics, University of Tartu, W.
Ostwaldi 1, Tartu 50411, Estonia}
\author{Rafael Ferraro}
\altaffiliation[]{ferraro@iafe.uba.ar} \affiliation{Instituto de
Astronom\'\i a y F\'\i sica del Espacio (IAFE, CONICET-UBA), Casilla
de Correo 67, Sucursal 28, 1428 Buenos Aires, Argentina.}
\affiliation{Departamento de F\'\i sica, Facultad de Ciencias
Exactas y Naturales, Universidad de Buenos Aires, Ciudad
Universitaria, Pabell\'on I, 1428 Buenos Aires, Argentina.}

\author{Alexey Golovnev}
\altaffiliation[]{agolovnev@yandex.ru} \affiliation{Centre for
Theoretical Physics, The British University in Egypt, 11837 El
Sherouk City, Egypt}

\author{Mar\'ia-Jos\'e Guzm\'an}
\altaffiliation[]{mjguzman@ut.ee} \affiliation{Laboratory of
Theoretical Physics, Institute of Physics, University of Tartu, W.
Ostwaldi 1, Tartu 50411, Estonia}

\date{\today}

\begin{abstract}
General relativity dynamics can be derived from different actions
--which depart from the Einstein-Hilbert action in boundary terms--
and for different choices of the dynamical variables. Among them,
the \textit{teleparallel equivalent of general relativity} is a
torsion-based theory for the tetrad field.  More general
torsion-based theories have been built in the last years, intending
to supersede general relativity. There are two current ways to
formulate such theories; one includes a spin-connection and the
other does not. We discuss the notion of Lorentz gauge invariance in
such theories, and give a simple but important proof that both
formulations are physically equivalent.

\end{abstract}

\maketitle

\section{Introduction}

\textit{Fernparallelismus} or teleparallelism is firstly known as an
attempt by Einstein to base a unified theory of the electromagnetic
and gravitational fields on the mathematical structure of absolute
teleparallelism. In this framework, spacetime is characterized by a
curvatureless connection together with a metric tensor field, both
defined in terms of a dynamical \textit{tetrad} or {\it vierbein}
field (a field of orthonormal tangent-space bases). This was an
episode in Einstein's research that lasted for three years from 1928
to 1931. Although the pertinent  mathematical structures had been
developed before by Cartan and Weitzenb\"{o}ck, they were introduced
just as mathematical concepts, and the search for field equations
for the tetrads was the novelty, efforts which eventually were
abandoned \cite{Sauer:2004hj}.

The concepts of teleparallelism were not further explored until
three decades later, when M{\o}ller \cite{Moller} revived Einstein's
original idea, but in order to find a tensorial complex for the
gravitational energy-momentum density. After this, the Lagrangian
formulation for teleparallel gravity was written by Pellegrini and
Plebianski \cite{Plebanski}. Later, Hayashi and Shirafuji
\cite{Hayashi} proposed the New General Relativity (NGR), a
teleparallel theory based on the curvatureless Weitzenb\"{o}ck
connection (which is linear in first derivatives of the tetrad). The
NGR Lagrangian combines the three quadratic invariants emerging from
the decomposition of the Weitzenb\"{o}ck torsion in its irreducible
parts. Three parameters enter the NGR Lagrangian, which can be fixed
to render NGR the {\it teleparallel equivalent} of general
relativity (TEGR). The equivalence between TEGR and GR  lies in the
fact that TEGR dynamics for the tetrad yields the GR dynamics for
the metric whenever the tetrad is thought as an {\it orthonormal}
basis in the tangent space, so linking the tetrad to the metric. The
orthonormality property is clearly invariant under {\it local}
Lorentz transformations of the tetrad. The equivalence between TEGR
and GR would suggest that the TEGR Lagrangian should exhibit such a
{\it gauge invariance}. Actually, TEGR Lagrangian is not invariant
but {\it pseudo-invariant} \cite{Sundermeyer}: a local Lorentz
transformation of the tetrad produces a boundary term in the TEGR
action, which does not affect the dynamics.

In the middle of the 2000's, Ferraro and Fiorini proposed a {\it
non-linear} extension of TEGR theory that consisted in replacing the
{\it torsion scalar} $\mathbb{T}$ in the Lagrangian with a
convenient function of it. For a Born-Infeld-like Lagrangian, they
have shown that the inflationary early-universe is a solution for
ordinary sources such as radiation or dust \cite{Ferraro:2006jd}.
This opened the study of the so called $f(\mathbb{T})$ theories of
modified teleparallel gravity \cite{Ferraro:2008ey,
Bengochea:2008gz, Linder}. We should notice that, although
pseudo-invariance works like a full invariance at the level of the
TEGR equations of motion, this feature is lost when passing to
$f(\mathbb{T})$ gravity. The divergence associated with the boundary
term now remains encapsulated within the argument of the function
$f(\mathbb{T})$ in the Lagrangian. Thus the equations of motion are
no longer invariant under local Lorentz transformations of the
tetrad (however, they remain invariant under global Lorentz
transformations). The loss of this gauge freedom implies that
$f(\mathbb{T})$ theories contain some degrees of freedom ({\it
d.o.f.}) associated with the local orientation of the tetrad field
(i.e., with the {\it parallelization} of the spacetime). How many
extra {\it d.o.f.} are involved in $f(\mathbb{T})$ and other
modified teleparallel gravities is a subject under study; but given
the non-linear background-dependent nature of its constraints, it
can be expected that the extra {\it d.o.f.} will depend on the point
in the phase space
\cite{Li:2011rn,Ferraro:2015,Ferraro:2018tpu,Ferraro:2018axk,Blagojevic:2020dyq,Blixt:2020ekl}.
As long as the matter couples just to the metric, or even the
Levi-Civita connection, the sole possibility of observing these
elusive extra {\it d.o.f.} would be via changes of gravitational
dynamics, but not at the level of the motion of matter in a given
spacetime.

The loss of the Lorentz gauge freedom can be avoided from the
beginning, by providing the TEGR Lagrangian with a divergence term
to guarantee its full local Lorentz invariance. However, this would
lead to a sort of  $f(R)$ theory (non-linear extension of the GR
Lagrangian), which would depart from the teleparallel foundations. A
different strategy consists of improving the Weitzenb\"{o}ck connection,
by replacing it with a more general one that is able to provide a
better behavior under local Lorentz transformations. In this way, we
endow the theory with a full Lorentz gauge invariance. For reasons
that will be clearer in the next Section, this strategy is called
{\it covariantization}.

In recent years we have been witnessing intense discussions about
the local Lorentz invariance in modified teleparallel theories of
gravity, such as $f(\mathbb{T})$ \cite{Golovnev:2017dox}. Some
authors object the covariantization procedure as inconsistent since
the components of the more general curvatureless connection cannot
be fixed by varying the covariantized action
\cite{Maluf:2018coz,Bejarano:2019fii}. Therefore the connection must
be chosen by resorting to some criterion, which is not very
different to directly adopting the Weitzenb\"{o}ck connection. Of
course, there might be some preferences based on symmetry
considerations, but those would often be not unique, neither
available in cases of no symmetry.

Some others view the local-Lorentz-covariant representations as the
only consistent approach \cite{Krssak:2015oua,Krssak:2018ywd}, by
assuming that the fundamental tetrad variable of teleparallel
gravity must be regarded as the frame an observer is free to choose,
and with the spin connection somehow related to inertial effects. In
such case, the Lorentz gauge freedom is frozen by parallelizing
according to some physical criterion. But if one thinks of the
tetrad as a geometric object to describe gravity, just a set of
orthonormal vectors without caring about observers, then such a
``physical'' criterion becomes meaningless. In any case, the extra
{\it d.o.f.} will enter into play and they turn out to be rather
unhealthy, at least for the standard $f(\mathbb{T})$ theory
\cite{Golovnev:2020nln,Golovnev:2020zpv} and NGR
\cite{Cheng:1988zg,BeltranJimenez:2019nns}.

All these disagreements seem odd, as there is evidence that the
covariant and the (original) pure tetrad approaches would be
equivalent \cite{Golovnev:2017dox,Golovnev:2021lki}.  Moreover, the
general curvatureless connection introduces trivial primary
constraints in the Hamiltonian analysis
\cite{Blixt:2019mkt,Blixt:2018znp}  which are evidently first-class
\cite{Golovnev:2021omn}. For instance, the so called preferred frame
effects or ``frame-dependent artifacts" are resolved only very
superficially \cite{Ren:2021uqb,Krssak:2015oua,Krssak:2018ywd}. In a
covariantized model one is allowed to choose absolutely any
available tetrad for a given metric, but at the price of taking an
appropriate spin connection. On the other hand, two different
solutions for the same metric ansatz of the pure tetrad approach
would also give two different solutions in the covariantized
version, in disguise of two different possible choices of the spin
connection for a given frame \cite{Golovnev:2021lki}.

It is not to say that the covariant approach has no physical
meaning, since for example, it naturally comes from taking
teleparallel gravity as a gauge theory of translations
\cite{Aldrovandi:2013wha}. Of course, this gives good hope for
formulating conserved quantities such as energy, for authors who
believe that these quantities should be well defined for gravity
despite that there is no fundamental symmetry which would make it
justified, so that in a really canonical way one gets rather a
holographic notion of conserved quantities, necessarily paying
attention to boundary terms \cite{Jimenez:2021nup}.

Leaving aside the controversies, the covariant methods can be used
for more convenience of doing calculations \cite{Golovnev:2021lki}.
However, what we will show in this article is that the pure-tetrad
formulation is precisely the same as simply taking the gauge
invariant variables in the covariant version, for any teleparallel
model which is globally Lorentz-invariant. This is yet another, very
simple and direct proof that the two versions have no physical
difference.

\section{Teleparallel and modified teleparallel gravity}
\label{sec:intro} In its standard formulation, teleparallel gravity
is built from the Weitzenb\"{o}ck connection, which is the connection
that makes the tetrad $e^b_{\nu}$, and its inverse $e_b^{\mu}$,
parallel transported as  $0=\nabla_\mu e^b_{\nu}=\partial_\mu
e^b_{\nu}-\Gamma^\alpha_{\mu\nu}e^b_\alpha$. Then, its components
are
\begin{equation}
\label{telecon}
\Gamma^{\alpha}_{\mu\nu}\equiv
e_a^{\alpha}\partial_{\mu}e^a_{\nu}\ .
\end{equation}
The Weitzenb\"{o}ck connection is compatible with the metric
$g_{\mu\nu}=\eta_{ab}e^a_{\mu}e^b_{\nu}$, or
$g_{\mu\nu}e_a^{\mu}e_b^{\nu}=\eta_{ab}$ (the basis
$\{\mathbf{e}_a\}$ is orthonormal). $\Gamma^{\alpha}_{\mu\nu}$
results to be a curvatureless connection whose torsion is
\begin{equation}
\label{torsion}
T^{\alpha}_{\hphantom{\alpha}\mu\nu}=\Gamma^{\alpha}_{\mu\nu}-\Gamma^{\alpha}_{\nu\mu}=
e_a^{\alpha}\, (\partial_{\mu}e^a_{\nu}-\partial_{\nu} e^a_{\mu})\ .
\end{equation}
The torsion scalar
\begin{equation}
\label{torscal} {\mathbb T} =\frac14
T_{\alpha\beta\mu}T^{\alpha\beta\mu}+\frac12
T_{\alpha\beta\mu}T^{\beta\alpha\mu}-T_{\mu}T^{\mu}\ ,
\end{equation}
where $T_{\mu}=T^{\alpha}_{\hphantom{\alpha}\alpha\mu}$ is the
torsion vector, is directly related to the Levi-Civita curvature
scalar $\overset{\circ}{R}$ of the metric $g_{\mu\nu}$ (that is the
GR Lagrangian),
\begin{equation}
\label{basrel} e \overset{\circ}{R} =-e {\mathbb T}+\partial_{\mu}(2
e T^{\mu}) ,
\end{equation}
where $e=\det e^a_\mu=(-\det g_{\mu\nu})^{1/2}$. On this basis, the
TEGR Lagrangian density is taken equal to $\pm e \mathbb{T}$, with
the sign depending on the chosen signature. Thus we obtain a TEGR
Lagrangian quadratic in first derivatives of the tetrad. The second
derivatives contained in the GR Lagrangian have been cornered in the
boundary term of Eq.~(\ref{basrel}), which is not relevant for the
dynamics.

In Eq.~(\ref{basrel}), the l.h.s. depends just on the metric, which
is invariant under {\it local} Lorentz transformations of the
tetrad,
\begin{equation}\label{Lor}
e^a_\mu\longrightarrow L^a_c(x) ~e^c_\mu\ \ \Longrightarrow \ \
g_{\mu\nu}\longrightarrow g_{\mu\nu} \ ,
\end{equation}
where $L^a_c$ is a matrix belonging to the Lorentz group (so it is
$\eta_{ab} L^a_c L^b_d=\eta_{cd}$). We remark that we are not
talking about the behavior under diffeomorphisms, since each term in
the r.h.s. of Eq.~(\ref{basrel}) is separately a scalar density.
Indeed the expression (\ref{telecon}) transforms as an affine
connection under coordinate changes, which implies that
$T^{\alpha}_{\hphantom{\alpha}\mu\nu}$ in Eq.~(\ref{torsion}) really
transforms as the components of a tensor. We are focusing our
analysis on the behavior of the r.h.s. of Eq.~(\ref{basrel}) under
local changes of the orientation of the tetrad. Of course, if the
l.h.s. of Eq.~(\ref{basrel}) possesses local Lorentz invariance,
then the r.h.s. will possess it too. However, separately each term
in the r.h.s. is only {\it globally} Lorentz invariant. This is
because the components of the torsion in Eq.~(\ref{torsion}) are
made of the components $\partial_{\mu}e^a_{\nu}-\partial_{\nu}
e^a_{\mu}$ of the exterior derivative of the tetrad $d{\mathbf
e}^a$, which changes as
\begin{equation}\label{exterior}
d{\mathbf e}^a\longrightarrow L^a_c~d{\mathbf
e}^c+d(L^a_c)\wedge{\mathbf e}^c \ .
\end{equation}
Only if the Lorentz transformation is global, then the 1-forms
$d(L^a_c)$ will be zero; thus $d{\mathbf e}^a$ will behave as a
2-form valued in the tangent space. In a more prosaic language, only
for global transformations, the label ``$a$" in $d{\mathbf e}^a$
will behave as a ``contravariant'' index under Lorentz
transformations of the tetrad, in the sense that $d{\mathbf e}^a$
will transform like ${\mathbf e}^a$.

The violation of the local Lorentz invariance does not show up in
the equations of motion of TEGR since its Lagrangian density differs
only by a boundary term $\mathbb{B}=\partial_\mu(2eT^\mu)$ from the
GR one. This makes TEGR {\it pseudo-invariant} under local Lorentz
transformations of the tetrad \cite{Ferraro:2020tqk}.
Pseudo-invariance is a rather low price to be paid for substituting
the GR Lagrangian, which contains second derivatives of the metric,
with the simpler TEGR Lagrangian that is built from  first
derivatives  of  the  tetrad. However, there are many
generalizations, such as NGR, $f(\mathbb{T})$ gravity, or even
higher derivative ones, which do violate the local Lorentz
invariance at the level of equations of motion. The most popular
higher derivative models are $f(\mathbb{T},\mathbb{B})$ ones, even
though they do not go much beyond the more usual modified gravities
and $f(\mathbb{T})$ gravity. Due to the basic relation
(\ref{basrel}), they can obviously be represented as simply
$f(\overset{\circ}{R} ,\mathbb{T})$, apparently inheriting all the
potential problems that  $f(\mathbb{T})$ could have
\cite{Golovnev:2020nln,Golovnev:2020zpv} .

\section{Lorentz-covariantization}
\label{sec:Lorentzcov}
The components of the torsion in
Eq.~(\ref{torsion}) will become invariant under local Lorentz
transformations of the tetrad if the ordinary exterior derivative
$d{\mathbf e}^a$ is {\it covariantized} by endowing it with a {\it
spin connection} term:
\begin{equation}
\label{covder} {\mathcal D} {\mathbf e}^a \equiv d{\mathbf e}^a
+\omega^a_{\hphantom{a} b}\wedge {\mathbf e}^b\ .
\end{equation}
The 1-forms $\omega^a_{\hphantom{a} b}$ must accompany the change of
the tetrad by transforming as components of a spin connection to
absorb the undesirable term in Eq.~(\ref{exterior}):
\begin{equation}
\label{covLor} {\mathbf e}^a\rightarrow L^a_c\, {\mathbf e}^c, \quad
\omega^a_{\hphantom{a} b} \rightarrow L^a_c\, \omega^c_{\hphantom{c}
d}\, (L^{-1})^d_b-(L^{-1})^c_b\ d(L^a_c)\ .
\end{equation}
Thus it turns out to be ${\mathcal D} {\mathbf e}^a \rightarrow
L^a_c(x)\, {\mathcal D} {\mathbf e}^c$.

When talking about invariance or covariance, it is very important to
precisely specify what kind of transformations are being considered,
since the invoked property is not only about the algebraic structure
of the involved group but also about how it acts. Let us introduce
the following two definitions:

\textbf{Definition 1:} If a teleparallel theory is Lorentz invariant
under the simultaneous transformations of the tetrad field and the
spin connection of Eq.~(\ref{covLor}), then it will be called
\textit{Lorentz invariant of type I}.

\textbf{Definition 2:} If a teleparallel theory is Lorentz invariant
under the transformation (\ref{Lor}) of the tetrad field (alone),
then it will be called \textit{Lorentz invariant of type II}.

\textbf{Remark:} Naturally, if a theory violates the Lorentz
invariance only at the boundary, the word ``pseudo'' is added to the
above definitions.

Any theory with an explicitly introduced spin connection is usually
Lorentz-invariant of type I, in the full meaning of that, without
the prefix ``pseudo". However, in case of  teleparallel models, the
type II is at best a ``pseudo-invariance" which gets broken by
almost every modification away from TEGR. On the other hand, given
the fulfillment of type I invariance, the transformations of type II
can equivalently be viewed as transformations of the spin connection
alone.

In its pure tetrad formulation, TEGR uses the Weitzenb\"{o}ck spin
connection $\omega^a_{\hphantom{a} b}=0$. But a teleparallel theory
covariantly formulated should be built from the (Lorentz invariant)
torsion $T^{\alpha}_{\hphantom{\alpha}\mu\nu}=
e_a^{\alpha}({\mathcal D} {\mathbf e}^a)_{\mu\nu}$
\footnote{${\mathbf T}^a={\mathcal D} {\mathbf e}^a$ is the 1st
Cartan's structure equation. It defines the relation between torsion
and connection.}. Since teleparallelism uses curvatureless
connections, so implying that gravity comes exclusively from the
torsion field, the spin connection should be chosen within the
family of curvatureless spin connections to which the Weitzenb\"{o}ck
connection belongs. The more general connection of this type can be
obtained by local-Lorentz transforming the Weitzenb\"{o}ck connection
(modulo possible global issues of cohomology type)
\begin{equation}
\label{spincon} \omega^a_{\hphantom{a} b}=-(\Lambda^{-1})^c_b\  d
\Lambda^a_c\ ,
\end{equation}
where the $\Lambda$'s are matrices belonging to the Lorentz group
\footnote{The basis $\{{\mathbf e}_a\}$ will no longer be parallel
transported if the Weitzenb\"{o}ck connection is replaced with
(\ref{spincon}), but the connection will be still metric. Anyway,
the ``parallelization'' can be always retrieved by passing to the
Weitzenb\"{o}ck connection through a local Lorentz transformation.}. We
use ``$\Lambda$'' instead of ``$L$'' because they are new variables
of the theory, characterizing the spin connection, while ``$L$''
indicates the local Lorentz transformations of the variables.

Now the (type I) simultaneous Lorentz transformation (\ref{covLor})
can be displayed as
\begin{equation}
\label{eLLor} e^a_\mu  \longrightarrow L^a_c (x)\, e^c_\mu\,, \quad
\Lambda^a_b \longrightarrow L^a_c (x)\, \Lambda^c_b.
\end{equation}

As explained above, any (modified) teleparallel model which is
globally Lorentz invariant in its pure tetrad formulation (every
model discussed in the literature we know) becomes locally Lorentz
invariant upon this covariantization procedure, though with respect
to the (type I) simultaneous transformation (\ref{covLor}).

For many popular models, it has been shown that the equation of
motion for the spin connection, which results from varying the
action with respect to the $\Lambda$'s variables, just reproduces
the antisymmetric part of the tetrad equations. Why this happens is
rather evident \cite{Golovnev:2017dox}: in the Lorentz-covariant
action, the antisymmetric variation of the tetrad gets precisely
compensated by variation of the spin connection (keeping it inside
the flat metric-compatible class).

Therefore, in the covariant version we can always choose the
$\omega^a_{\hphantom{a} b}=0$ gauge, even right inside the action
which otherwise is not always a harmless choice to do, which brings
us back to the pure tetrad formalism.

Moreover, any gauge choice does not influence the value of the
torsion tensor, therefore it is not only that this choice does not
influence the physical contents of equations of motion, but it does
not change the global quantities, like the full value of the action,
either. If we found some solution in the covariant version, we can
choose a gauge and make it a pure-tetrad solution, with the same
metric and the same torsion tensor.

\section{Lorentz gauge-invariant variables}
\label{sec:Lorentzvar}
Now, when we know that any globally
Lorentz-invariant modified teleparallel model can be covariantized
by replacing the partial derivatives of the tetrad for the
Lorentz-covariant ones, let us make a simple rewriting of the
covariant derivative (\ref{covder}) with the connection
(\ref{spincon}):
\begin{eqnarray}
\label{trans} ({\mathcal D} {\mathbf e}^a)_{\mu\nu}&=&
2\;\partial_{[\mu} e^a_{\nu]}-2\;(\Lambda^{-1})^c_b\,
(\partial_{[\mu}\Lambda^a_c)\, e^b_{\nu]}\cr\cr
&=&2\;\partial_{[\mu} e^a_{\nu]}+2\;\Lambda^a_c\,
(\partial_{[\mu}(\Lambda^{-1})^c_b)\, e^b_{\nu]}\cr\cr
&=&\Lambda^a_c\ 2\left[(\Lambda^{-1})^c_b\, \partial_{[\mu}
e^b_{\nu]}+(\partial_{[\mu}(\Lambda^{-1})^c_b)\,  e^b_{\nu]}\right]\
\ \cr\cr &=&\Lambda^a_c\ 2\, \partial_{[\mu} {\tilde e}^{\; c}_{\;
\nu]}\ ,
\end{eqnarray}
which means that ${\mathcal D} {\mathbf e}^a=\Lambda^a_c\ d {\tilde
{\mathbf e}}^{c}$, where
\begin{equation}
\label{invE} {\tilde {\mathbf e}}^{ c}\equiv (\Lambda^{-1})^c_b\
{\mathbf e}^b
\end{equation}
is a Lorentz-invariant quantity which does not change at all under
the simultaneous local Lorentz transformations (\ref{eLLor}). Thus
the Lorentz-covariant torsion tensor is
\begin{equation}\label{covtor}
T^{\alpha}_{\hphantom{\alpha}\mu\nu}=e_a^\alpha\, ({\mathcal D}
{\mathbf e}^a)_{\mu\nu}={\tilde e}_{\; a}^{\; \alpha}\;
(\partial_{\mu} {\tilde e}^{\; a}_{\; \nu}-\partial_{\nu} {\tilde
e}^{\; a}_{\; \mu})\ ,
\end{equation}
which is the torsion for the Weitzenb\"{o}ck connection associated with
${\tilde {\mathbf e}}^a$. Then, the covariant formulation is
dynamically equivalent to a pure tetrad formulation. The only
dynamical object of the covariant formulation is ${\tilde {\mathbf
e}}^a$:

\begin{equation}
\mathcal{L}^{\mathrm{covariant}}(e,\Lambda)=\mathcal{L}^{\mathrm{pure\
tetrad}}(\tilde e)\ , \label{redef}
\end{equation}
and therefore the locally Lorentz-invariant variables (\ref{invE})
simply satisfy the equations of motion of the pure tetrad version of
the model. Thus, conclusions made in the pure tetrad formulation
also apply to the covariant formulation, and it is evident that the
spin connection contains purely gauge degrees of freedom.

The formula (\ref{redef}) is valid for any modified teleparallel
model which can be written in terms of the metric (and its
Levi-Civita connection) and the spacetime components of the torsion
tensor. Indeed, the metric is obviously invariant under any kind of
the Lorentz transformations, be it type I or type II, while for the
torsion tensor we have
$$T^{\alpha}_{\hphantom{\alpha}\mu\nu}(e,\omega(\Lambda))=T^{\alpha}_{\hphantom{\alpha}\mu\nu}(\tilde e,0).$$
Therefore, it includes $f(\mathbb T)$ models as well as NGR and its
non-linear generalizations, and even models of $f(\mathbb T, \mathbb
B)$ and many other types.

What we have shown here is that taking the $\omega^a_{\hphantom{a}
b}=0$ gauge is equivalent to a simple change of the variables. Very
similarly, a Lagrangian of the form $$\mathcal{L}(\phi,\psi)=\frac12
\partial_{\mu}(\phi-\psi)\partial^{\mu}(\phi-\psi)$$ can be
transformed to $\mathcal{L}(\chi,\psi)=\frac12
(\partial_{\mu}\chi)(\partial^{\mu}\chi)$ by a simple change of
variable $\phi\to\chi=\phi-\psi$ which removes its dependence on
$\psi$. It is nothing but rewriting the Lagrangian in
gauge-invariant variables.

A source of concern might be eligibility of restricting to
gauge-invariant variables only, inside the action. For example,
using the vector potential is important for the action principle of
electrodynamics. In fact, its gauge-invariant quantities
$F_{\mu\nu}=\partial_{\mu}A_{\nu}-\partial_{\nu}A_{\mu}$ depend on
derivatives of the fundamental variables. The variation of
$\mathcal{L}=-\frac14 F_{\mu\nu}F^{\mu\nu}$ with respect to
$F_{\mu\nu}$ would give the trivial equation $F_{\mu\nu}=0$ while
its variation with respect to $A_{\nu}$ gives the correct result of
$\overset{\circ}{\triangledown}_{\mu} F^{\mu\nu}=0$.

The reason is that the condition of $\delta A$ vanishing at infinity
is stronger than the same condition for $\delta F$. This is not
about a precise type of asymptotic behavior, the same is true for
vanishing everywhere outside a big enough ball. For example,
vanishing of an otherwise arbitrary function $x(t)$ both in the past
and in the future means also that $\dot x(t)$ integrates to zero,
what gives a non-trivial condition on admissible functions $\dot
x(t)$ even if the boundary contributions are totally neglected. In
other words, the usual variational principle in terms of $A_{\mu}$
imposes extra restrictions on the allowed variations of
$F_{\mu\nu}$.

The stronger condition on admissible variations means that the
action must be stationary with respect to a smaller class of
variations. It makes the corresponding equations admit more
solutions. Note also that the vanishing field strength is of course
a particular case of the divergenceless one. This is a common effect
which plays, for example, an important role in understanding mimetic
gravity models \cite{Golovnev:2013jxa}.

However, this is not the case with the choice of variables presented
in this work; covariant and pure-tetrad formulations are  equivalent
at the level of the action too. This is because the gauge-invariant
variables (\ref{invE}) do not depend on any derivatives of the
fundamental quantities; we have presented a purely algebraic
relation which imposes no restriction on the class of variations of
gauge-invariant variables.

There are two different sides of the traditional understanding of
the role of spin connection \cite{Krssak:2018ywd}. One assumes a
particular spin connection for a given tetrad, which removes
inertial effects. This assumption that one could objectively
separate gravity from inertia goes against all experimental
evidence. And it is related to the picture of (fictional) global
translations being gauged, and requires some, necessarily voluntary,
choice of the reference tetrad. By now, it is known that the
standard recipes \cite{Krssak:2018ywd} of determining the spin
connection can give non-unique results \cite{Emtsova:2021ehh}, and
beyond the simplest cases they might even easily get wrong, with
equations of motion not being satisfied if not in TEGR
\cite{Bahamonde:2020snl}.

Secondly, there are opinions in the literature concerned about
making the action and other global quantities finite, and it has
been sometimes claimed that the spin connection plays a regularizing
role. However, our change of variables is performed directly at the
level of all the fundamental geometrical quantities, and does not
involve neglecting any boundary term. It means that the action, or
some conserved quantities, can be regularized in the pure tetrad
approach with no less success than in the covariant one. And in
fact, if there is a finite-action covariant solution, we can always
make a type I (simultaneous) Lorentz transformation which brings
this particular solution, preserving all its physical properties, to
the zero spin connection case.

To state it once more, there are different types of the local
Lorentz (pseudo-)invariance in the case of TEGR. One way is to write
the action fully in terms of the Lorentz-covariant derivatives. Then
we have the type I invariance, in practically any model we might
think of, in these terms. On the other hand, if we think of the
tetrad as four vectors composing the fundamental degrees of freedom,
then there is no reason to want this full-fledged local Lorentz
invariance.

However, even the pure-tetrad action of TEGR appears to possess some
invariance which goes beyond the explicitly maintained
diffeomorphism invariance. This is the type II Lorentz
pseudo-invariance. And this is the invariance that gets lost in most
generalizations. Even though it is different from what most experts
in general metric-affine gravity are used to call local Lorentz
invariance, it is algebraically related with the same group, the
Lorentz group. Note also that the influence of all these symmetries
of the action can be seen in the Hamiltonian analysis at the level
of the primary constraints \cite{Golovnev:2021omn}.

Of course, what we get in models of $f(T)$ gravity is non-trivial
dependence on which tetrad to choose. But this is not an obstacle,
since it grants a new model with more degrees of freedom. These are
represented by the tetrad, a fundamental variable which carries more
information than what the metric does. The covariantization
procedure is equivalent to formally rewriting the very same model in
locally Lorentz invariant terms. This is done by simply sharing the
extra modes with the newly introduced variables, the components of
the spin connection. However it does not get rid of the very fact
that the model does have some new dynamical content which a metric
alone cannot provide. Getting something more on top of the metric is
the essence of the local Lorentz symmetry (type II) violation, and
the covariant version simply rewrites it in different terms.

\section{Conclusions}
\label{sec:Conclusions}
We have proven that the covariant and
pure-tetrad formulations of teleparallel theories are fully
equivalent, as long as matter is not coupled to the {\it Lorentzian}
extra degrees of freedom. Due to the nature of the (curvatureless
and metric) Weitzenb\"{o}ck connection (\ref{telecon}), its torsion can
be covariantized by replacing the tetrad ${\mathbf e}^{a}$ with the
``gauge invariant'' tetrad ${\tilde {\mathbf e}}^{a}$ (see
Eq.~(\ref{covtor})), which is equivalent to splitting the original
tetrad into two factors (see Eq.~(\ref{invE})). This splitting
introduces six new variables -- the Lorentz matrices $\Lambda^a_b$
-- deprived of independent dynamics; they are clearly spurious
variables that will have an impact on the constraint algebra. Since
the covariantized torsion is the Weitzenb\"{o}ck torsion of the tetrad
${\tilde {\mathbf e}}^{a}$, then ${\tilde {\mathbf e}}^{a}$
satisfies the same dynamics as the original variables ${\mathbf
e}^{a}$. Therefore, although the covariant formulation is Lorentz
invariant of type I, it is not dynamically different from the pure
tetrad formulation.

When the Lorentz invariance of type II is broken in the pure tetrad
formulation, it gives rise to new dynamics. This dynamics also
appears in the covariant formulation, because the Lorentz type II
breaking is also present. Certainly, a pure-tetrad Lorentz rotation
is no longer an obvious symmetry once non-zero spin connection terms
are added. Therefore, what is broken by generalizations of the
covariant TEGR would rather be a ``type $\tilde{\mathrm{II}}$"
invariance, which represents local Lorentz rotations of our
Lorentz-invariant variables (\ref{invE}).

The results here exposed are rooted in the fact that Weitzenb\"{o}ck
connection (\ref{telecon}) has already the most general form for a
curvatureless connection. From this perspective, there is a
one-to-one relation between tetrads and curvatureless
tetrad-compatible connections (up to global linear transformations
and possible topological obstructions). This is analogous to the
(metric based) GR formulation, where there exists a one-to-one
relation between metrics and torsionless metric-compatible
(Levi-Civita) connections. However, since the relation tetrad-metric
is not one-to-one but is subject to a local Lorentz invariance, the
teleparallel formulations are expected to be endowed with such
invariance. This expectation seems to come from considering the
metric as more fundamental than the tetrad. However, such invariance
is not mandatory for building a dynamical theory of tetrads, since
both the gravity Lagrangian and the coupling matter-gravity can be
written only in terms of tetrads (without prejudice to those matter
Lagrangians that in fact exhibit a local Lorentz invariance).

Our findings are applicable to the entire class of teleparallel
theories built from the torsion tensor, which include TEGR,
$f(\mathbb T)$, NGR, $f(T_{ax},T_{vec},T_{ten})$,
$f(\mathbb{T},\mathbb{B})$, teleparallel Horndeski gravity, among
others.

\textbf{Acknowledgments:} The authors thank Laur J\"{a}rv for helpful
feedback on a previous version of this manuscript. DB and MJG have
been supported by the European Regional Development Fund CoE program
TK133 ``The Dark Side of the Universe''. MJG was funded by the
Estonian Research Council grant MOBJD622. RF was supported by
Consejo Nacional de Investigaciones Cient\'{\i}ficas y T\'{e}cnicas (CONICET)
and Universidad de Buenos Aires.

\end{document}